\documentclass[prd,twocolumn]{revtex4}
\usepackage{dcolumn}
\usepackage{multirow}
\usepackage{graphicx}
\usepackage{amssymb}
\usepackage{bm}
\usepackage{hyperref}
\usepackage{epstopdf}
\usepackage{color}
\usepackage{mathrsfs}
\usepackage{amsmath,amssymb,amsthm}
\usepackage{rotating}
\usepackage{sverb, longtable}
\usepackage{subfigure}

\usepackage[graphicx]{realboxes}
\usepackage{adjustbox}
\begin{document}

\title{Testing the Hubble law with Pantheon+}
\author{Deng Wang}
\email{cstar@nao.cas.cn}
\affiliation{National Astronomical Observatories, Chinese Academy of Sciences, Beijing, 100012, China}
\begin{abstract}
The Hubble law (HL) governs the low-redshift (low-z) evolution of the distance of an object. However, there is a lack of an investigation of its validity and effective radius for a long time, since the low-z background data with a high precision is scarce. The latest Type Ia supernovae sample Pantheon+ having a significant increase of low-z data provides an excellent opportunity to test the HL. We propose a generalized HL and implement the first modern test of the HL with Pantheon+. We obtain the constraint on the deviation parameter $\alpha=1.00118\pm0.00044$, confirm the validity of linear HL with a $0.04\%$ precision and give the transition redshift $z_t=0.03$ and luminosity distance $D_{L,t}=123.13\pm1.75$ Mpc, which means that HL holds when $z<0.03$ and breaks down at a distance of $D_L>123.13$ Mpc. Comparing the ability of Type Ia supernovae and HII galaxies in testing the HL, we stress the uniqueness and strong power of Type Ia supernovae in probing the low-z physics.        
\end{abstract}
\maketitle

{\it Introduction}. In 1922, Friedman first derived the so-called Friedmann equations characterizing the expanding universe in the framework of general relativity \cite{Friedman:1922kd}. After five years, Lema\^{i}tre \cite{Lemaitre:1927} independently derived the solution of cosmic expansion from gravitational field equations, observed the proportionality between the distance to a celestial body and the recessional velocity of this body, and estimated the proportionality coefficient to be $625$ km s$^{-1}$ Mpc$^{-1}$. Two years later, Hubble \cite{Hubble:1929ig} confirmed the existence of cosmic expansion using observations, gave a more accurate value of $500$ km s$^{-1}$ Mpc$^{-1}$ to this coefficient, and derived the following well-known HL
\begin{equation}
v=H_0D_L,   \label{1}
\end{equation}  
where $v$ is the recessional velocity of an object, $H_0$ is the Hubble constant and $D_L$ is the luminosity distance of the same object. It is worth noting that Hubble obtained this relationship by deriving the recession velocities of objects from their redshifts, many of which were earlier measured and related to velocities by Slipher in 1917 \cite{Slipher:1917}. After making a simple Taylor series expansion, the HL can be expressed as      
 
\begin{figure}
	\centering
	\includegraphics[scale=0.5]{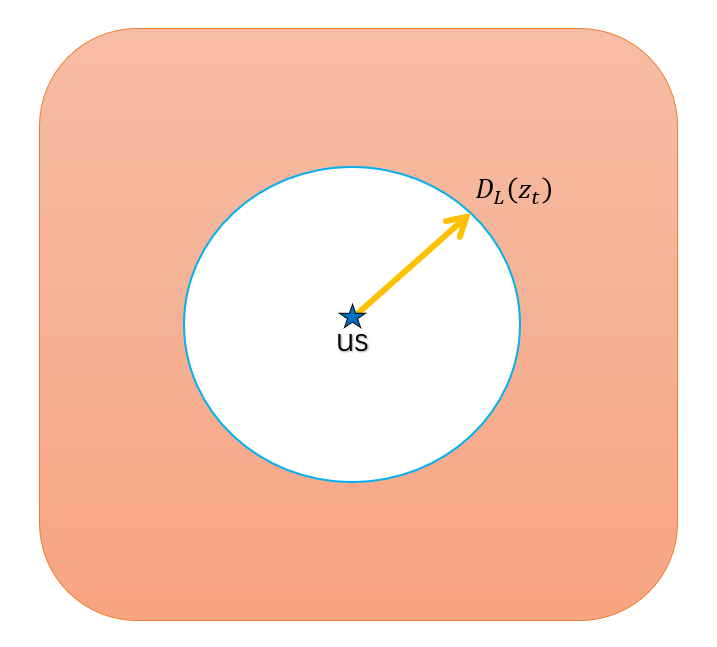}
	\caption{The schematic plot for the effective range of the HL. Below the transition redshift $z_t$, the HL $cz=H_0D_L$ holds but it breaks down beyond $z_t$.}\label{f1}
\end{figure}

\begin{figure*}
	\centering
	\includegraphics[scale=0.52]{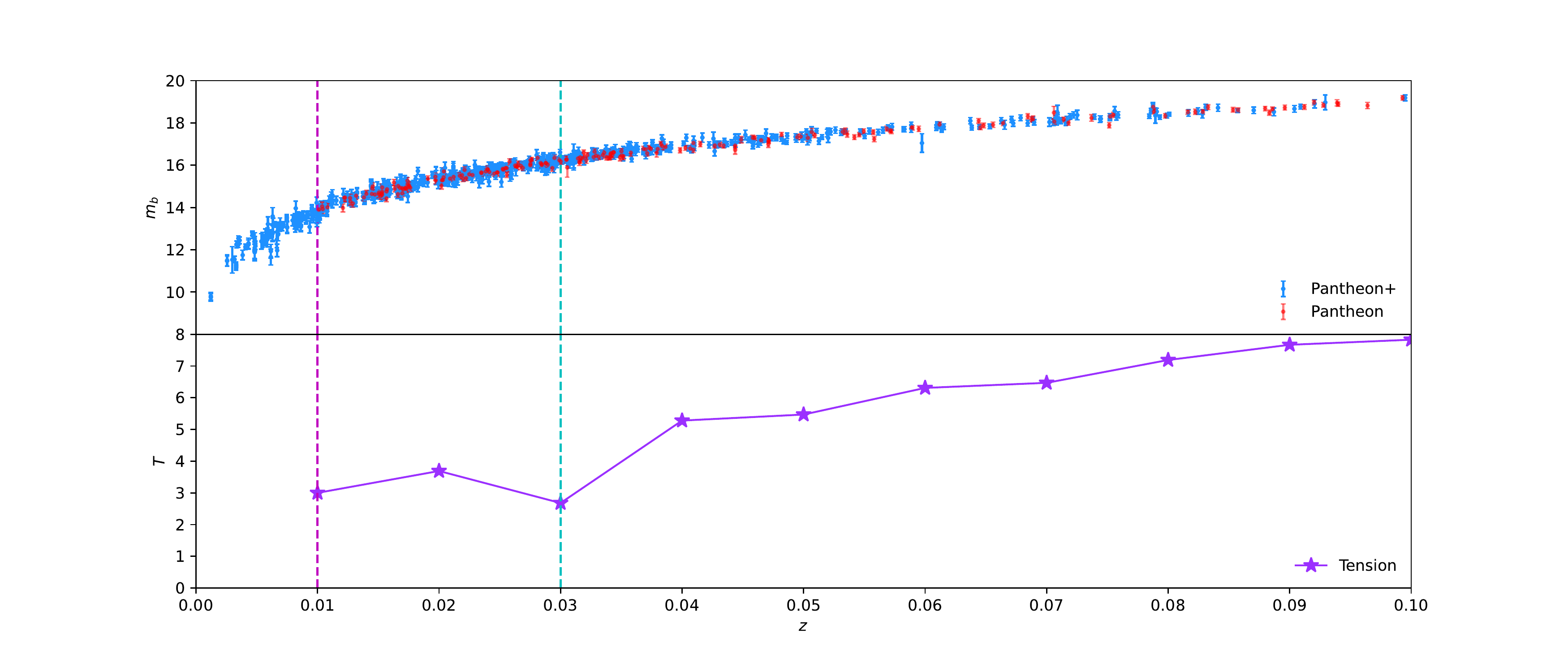}
	\caption{{\it Upper panel}. The redshift ($z$)-apparent magnitude ($m_b$) diagram of Pantheon+ (blue) and Pantheon (red). It is clear to see that Pantheon+ has more low-z SNe Ia and Pantheon has no data point in the range $z<0.01$. {\it Lower panel}. The evolution of the extent of the deviation from the standard HL prediction $\alpha=1$ over $z$. The dashed lines denote the redshift $z=0.01$ (magenta) and $z=0.03$ (cyan), respectively. }\label{f2}
\end{figure*}

\begin{table*}[!t]
	\renewcommand\arraystretch{1.5}
	\caption{The $1\,\sigma$ (68\%) constraining results of the generalized HL in different low-z redshift ranges. The number of SNe Ia and the extent of the deviation from $\alpha=1$ in each range is shown in the second and fourth columns, respectively.}
	\setlength{\tabcolsep}{12mm}{
		\begin{tabular} { l |c| c | c }
			\hline
			\hline
			
			Data range       & Number        & $\alpha$      & Tension                            \\
			\hline
			$z<0.01$         &111          &$1.00330\pm0.00110$    &$3.00\,\sigma$    \\
			\hline
			$z<0.02$         &265            &$1.00251\pm0.00068$     &$3.69\,\sigma$     \\
			\hline
			$z<0.03$         &468            &$1.00118\pm0.00044$     &$2.68\,\sigma$     \\
			\hline
			\hline
			$z<0.04$         &590          &$1.00190\pm0.00036$     &$5.28\,\sigma$     \\
			\hline		    
			$z<0.05$         &645             &$1.00186\pm0.00034$    &$5.47\,\sigma$     \\
			\hline		    
			$z<0.06$         &678           &$1.00202\pm0.00032$     &$6.31\,\sigma$     \\
			\hline		    
			$z<0.07$         &697      &$1.00207\pm0.00032$     &$6.47\,\sigma$     \\
			\hline		                                
			$z<0.08$         &727       &$1.00223\pm0.00031$     &$7.19\,\sigma$     \\
			\hline
			$z<0.09$         &736      &$1.00230\pm0.00030$    &$7.67\,\sigma$     \\
			\hline
			$z<0.1$         &741    &$1.00235\pm0.00030$     &$7.83\,\sigma$     \\
			\hline  
			\hline
			$z<0.2$         &948     &$1.00468\pm0.00024$     &$19.50\,\sigma$     \\
			\hline
			$z<0.3$         &1207     &$1.00688\pm0.00020$     &$34.40\,\sigma$     \\
			\hline
			\hline
		\end{tabular}
		\label{t1}}
\end{table*}

\begin{equation}
cz=H_0D_L,   \label{2}
\end{equation}
where $c$ and $z$ are the speed of light and the redshift of an observed object, respectively. This equation is a low-z approximate version of Eq.(1). At low redshifts, one can easily use it to derive the accurate redshift of an object when knowing its luminosity distance, and vice versa. However, in practice, with more and more high precision low-z observations, there are two problems occurs: (i) whether is actually HL valid for low-z observations at all? (ii) if HL is valid for low-z data, what is its effective radius? Koranyi and Strauss \cite{Koranyi:1997} investigated correctly the former problem by using the IRAS 1.2 Jy Redshift Survey in 1997, and confirmed the validity of HL. However, their data quality is very limited and they assume a large cosmic expansion rate $H_0=100$ km s$^{-1}$ Mpc$^{-1}$, which affects the correctness of their conclusions, to implement their analysis. Although HL as one of the milestones of modern cosmology tells us the fact that the universe is during an expanding phase, the validity of HL must be precisely tested. Unfortunately, so far, there is still a lack of an accurate test of HL with modern astronomical observations. Since we do not have enough data points covering the whole low-z range in the past several decades, the latter problem can not be solved well. Due to the rapid process of modern Type Ia supernovae (SNe Ia) observations, we can not only implement an accurate test of HL, but also determine well the luminosity distance $D_L(z_t)$ at the transition redshift $z_t$, beyond which $cz=H_0D_L$ will break down (see Fig.\ref{f1} for a better illustration). Particularly, the recent release of the largest SNe Ia sample called Pantheon+ \cite{Scolnic:2021amr,Brout:2022vxf}, which shows a significant increase of low-z data relative to the original Pantheon sample \cite{Pan-STARRS1:2017jku}, lets us have enough high-precision data points in the redshift range $z<0.1$ to test the HL. As a consequence, we implement the fist modern test of HL using the Pantheon+ data in this study. We confirm the validity of linear HL with a $0.04\%$ precision find that the HL holds when $z<0.03$ and breaks down at a distance of $D_L>123.13$ Mpc.     


\begin{figure*}
	\centering
	\includegraphics[scale=0.43]{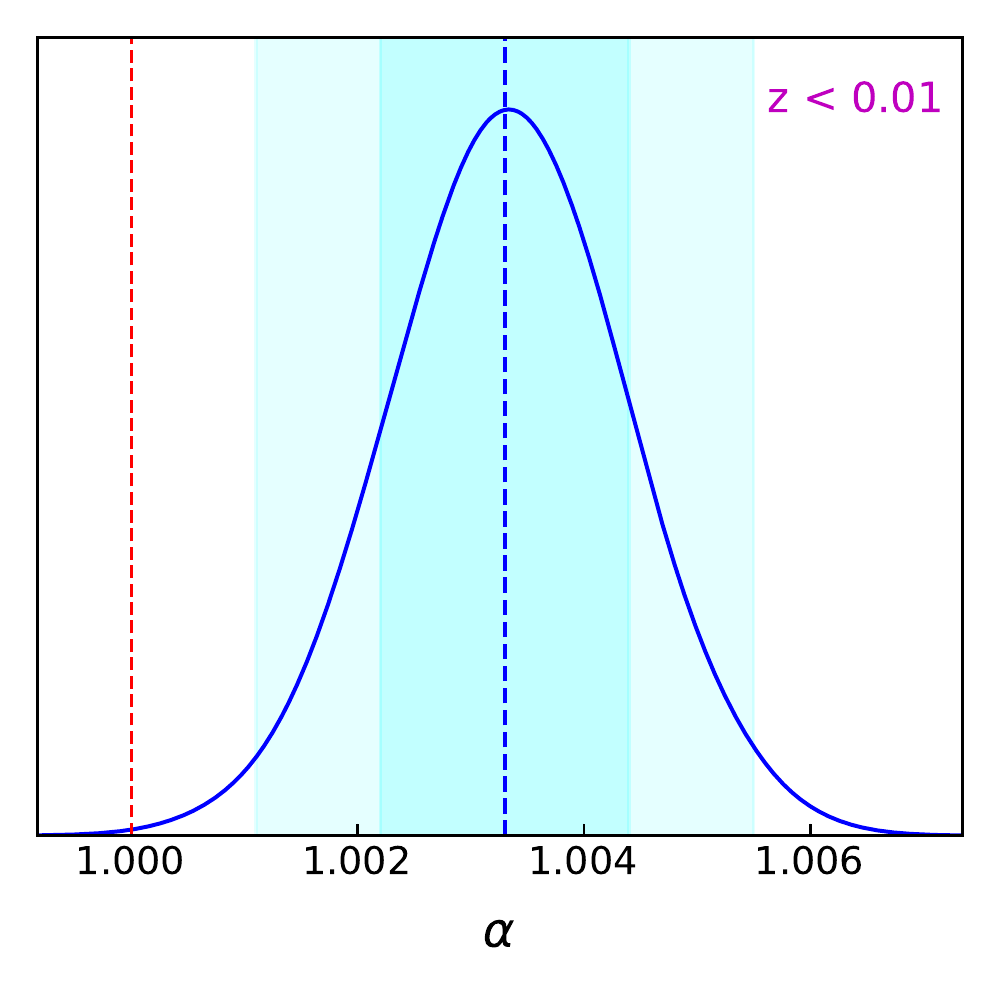}
	\includegraphics[scale=0.43]{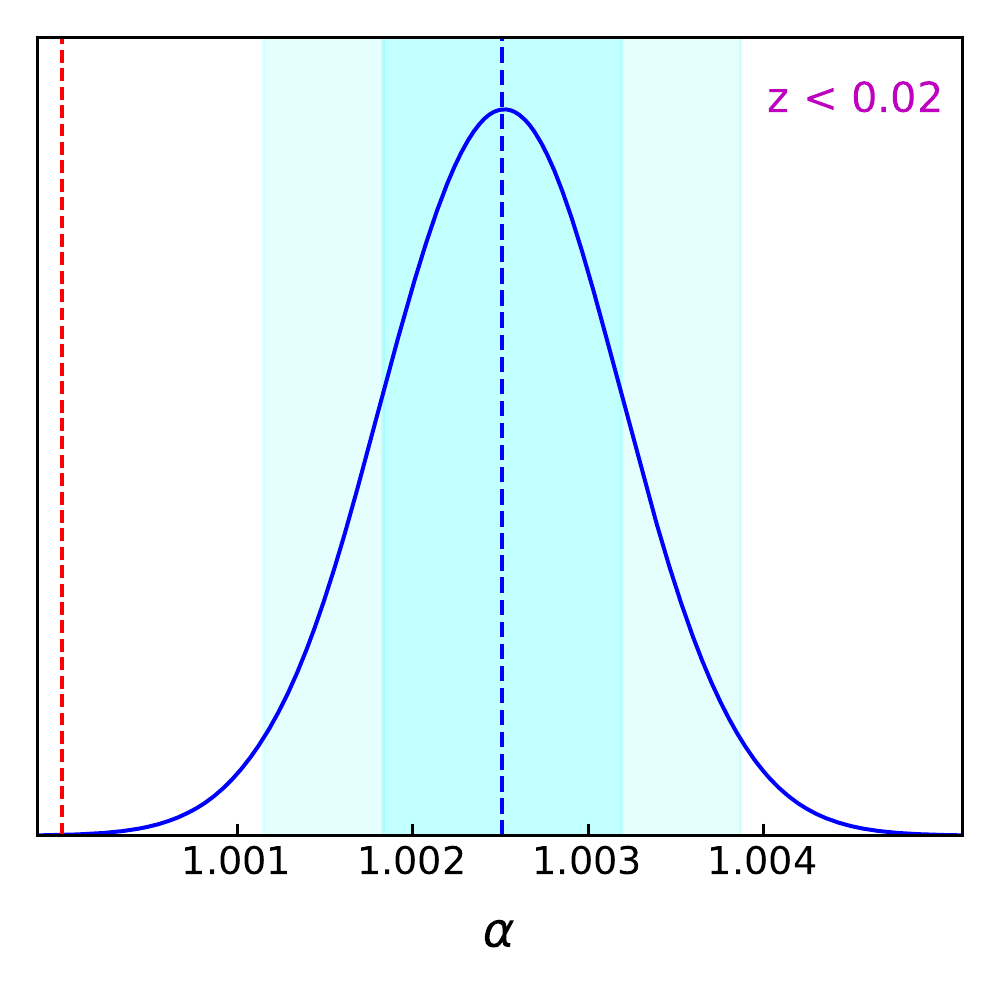}
	\includegraphics[scale=0.43]{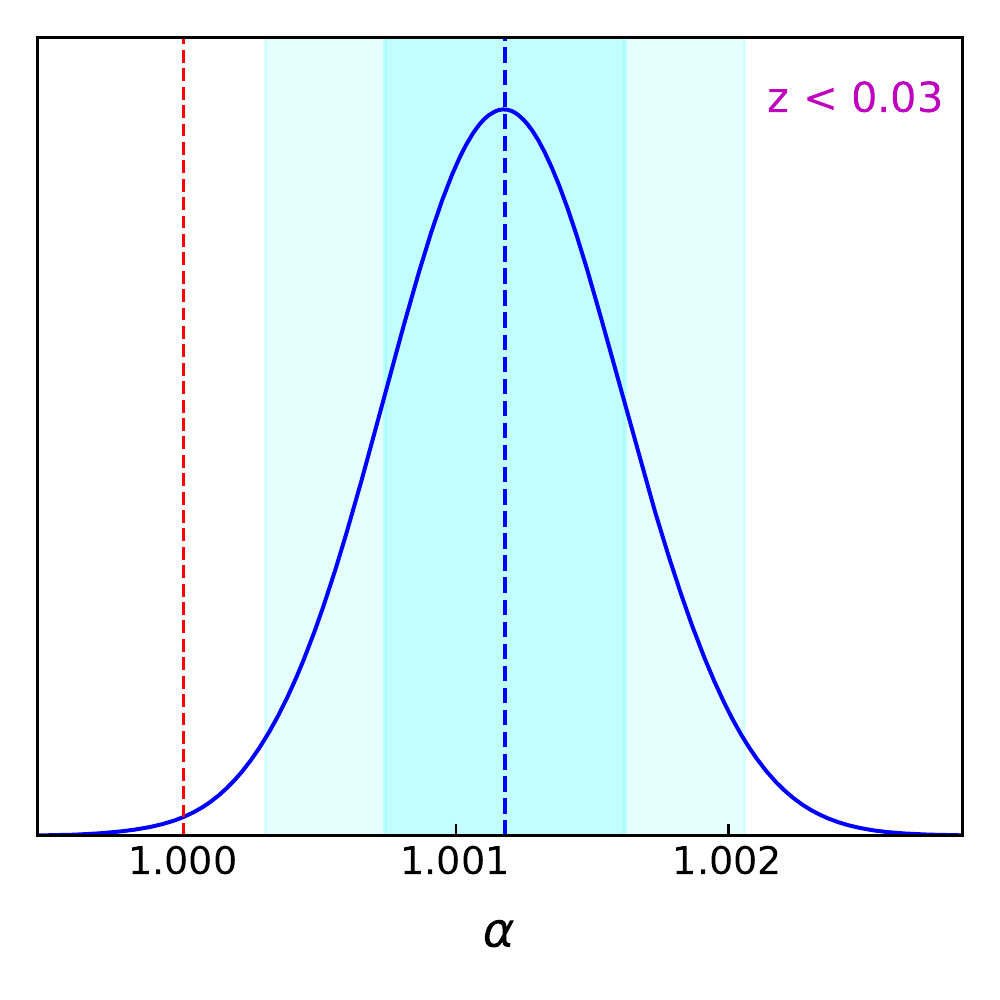}
	\includegraphics[scale=0.43]{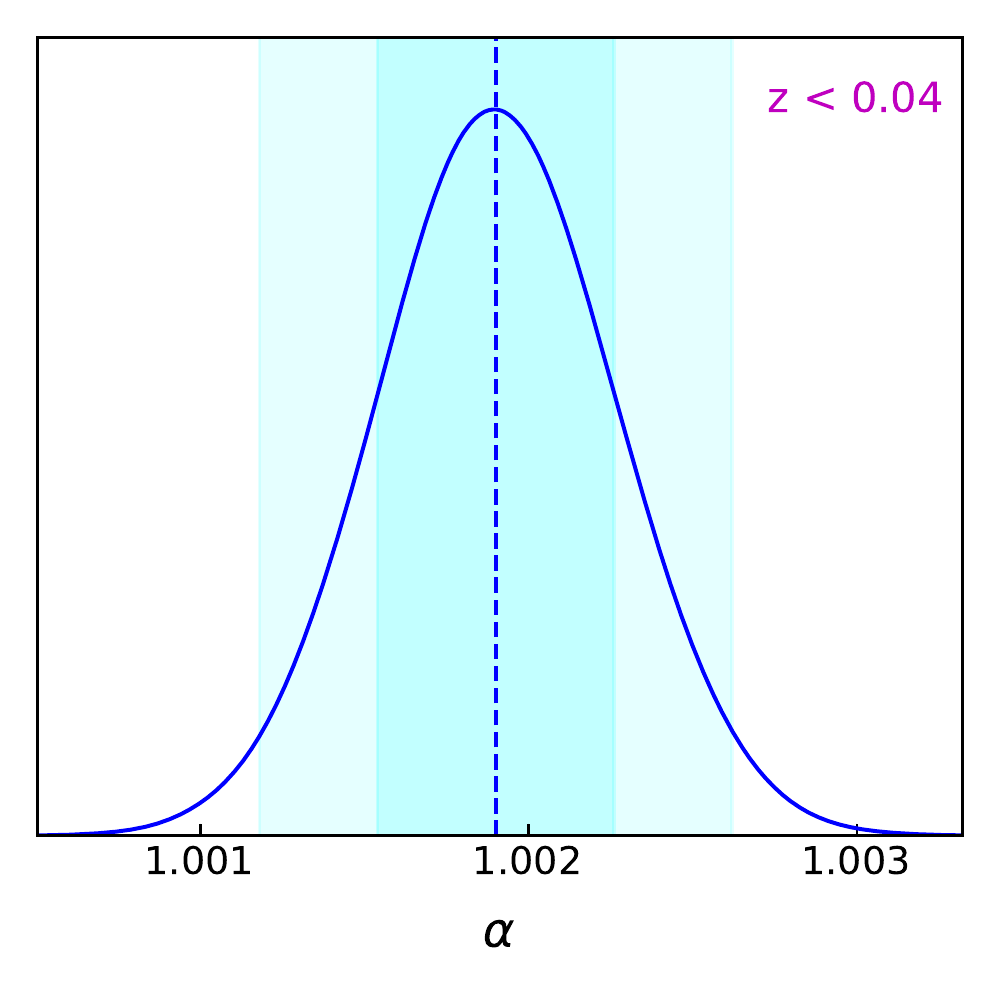}
	\includegraphics[scale=0.43]{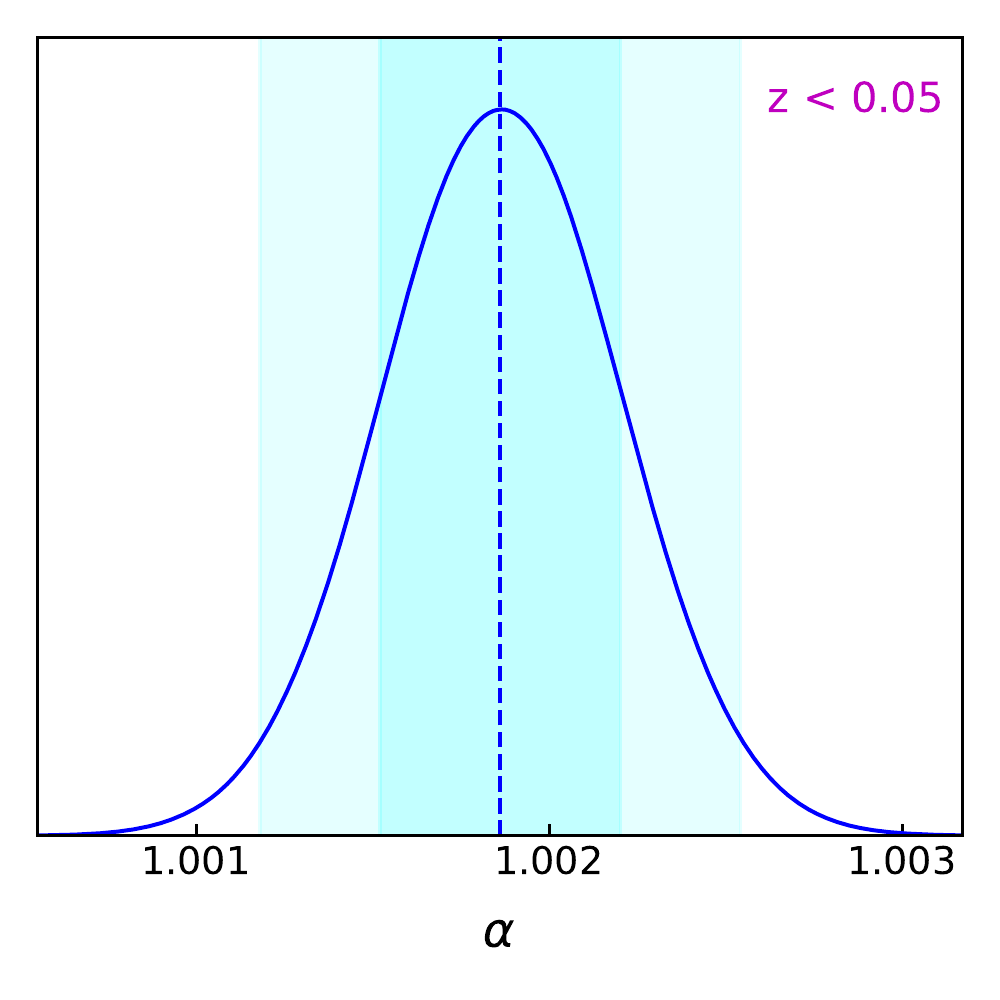}
	\includegraphics[scale=0.43]{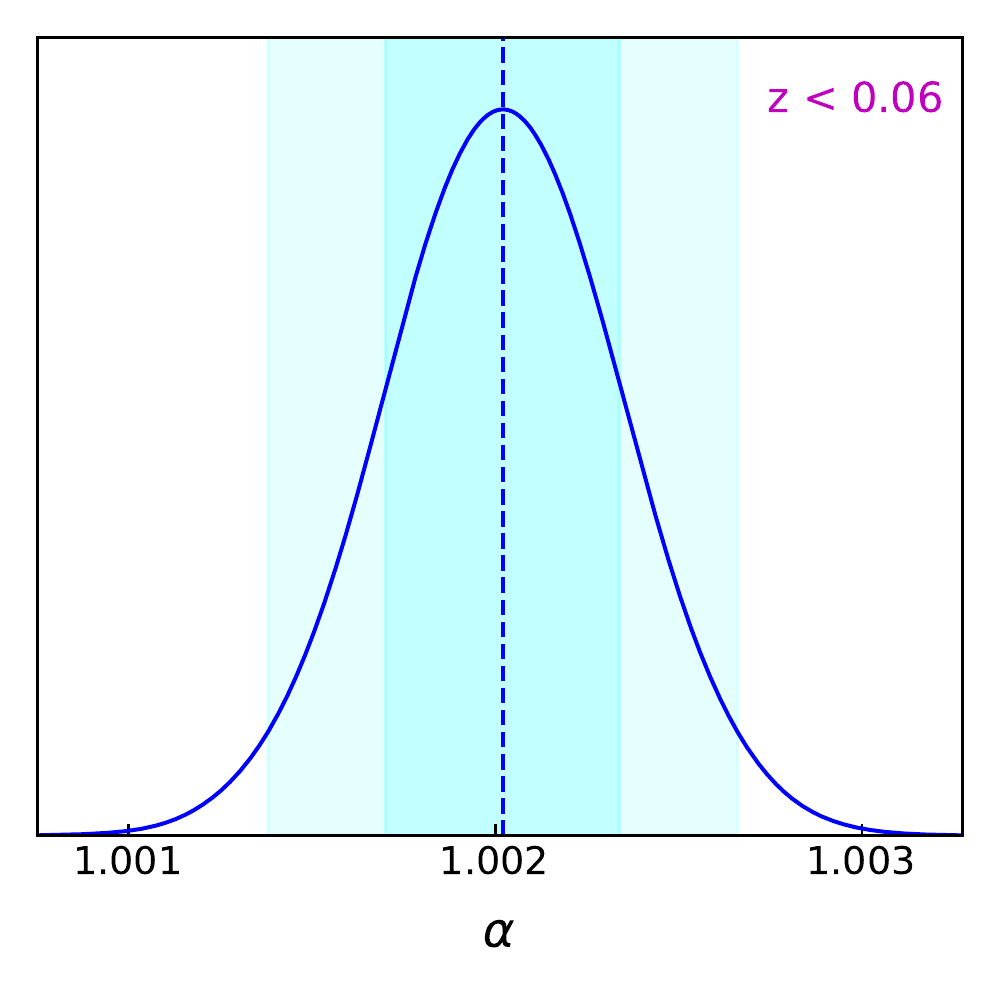}
	\includegraphics[scale=0.43]{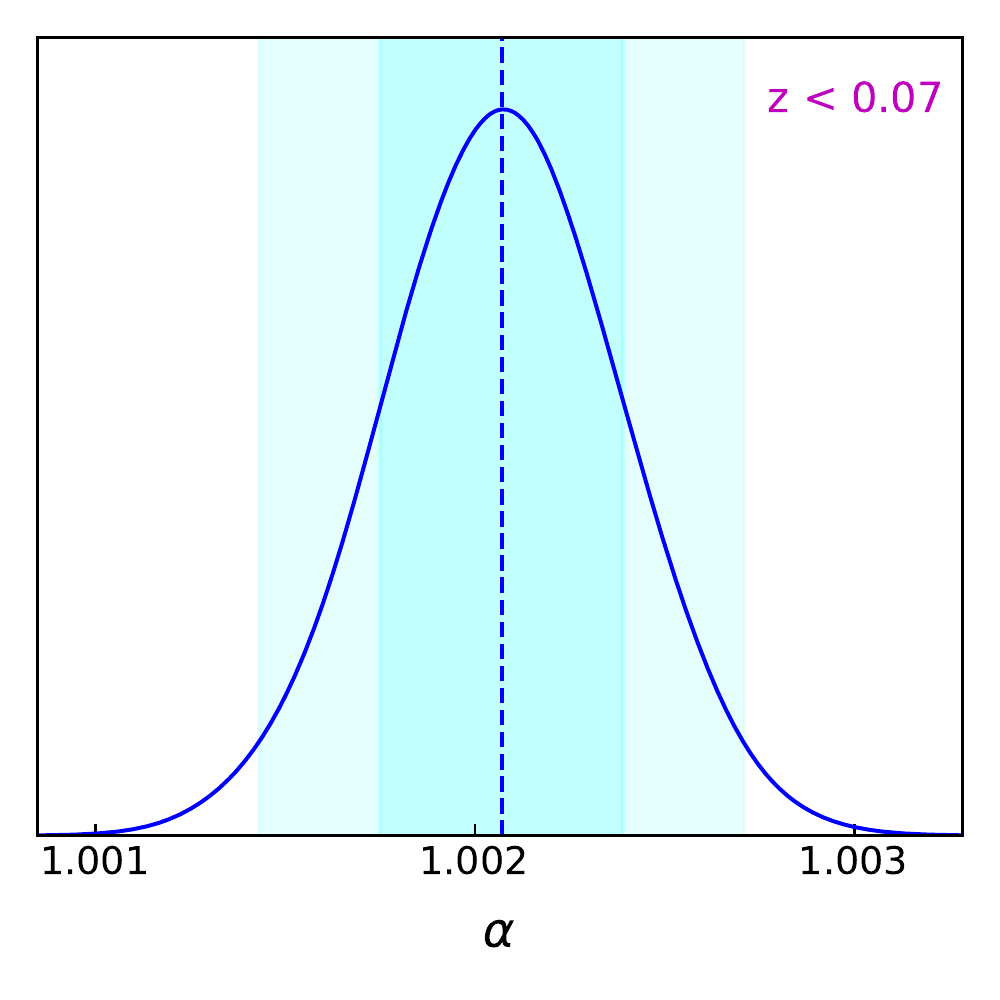}
	\includegraphics[scale=0.43]{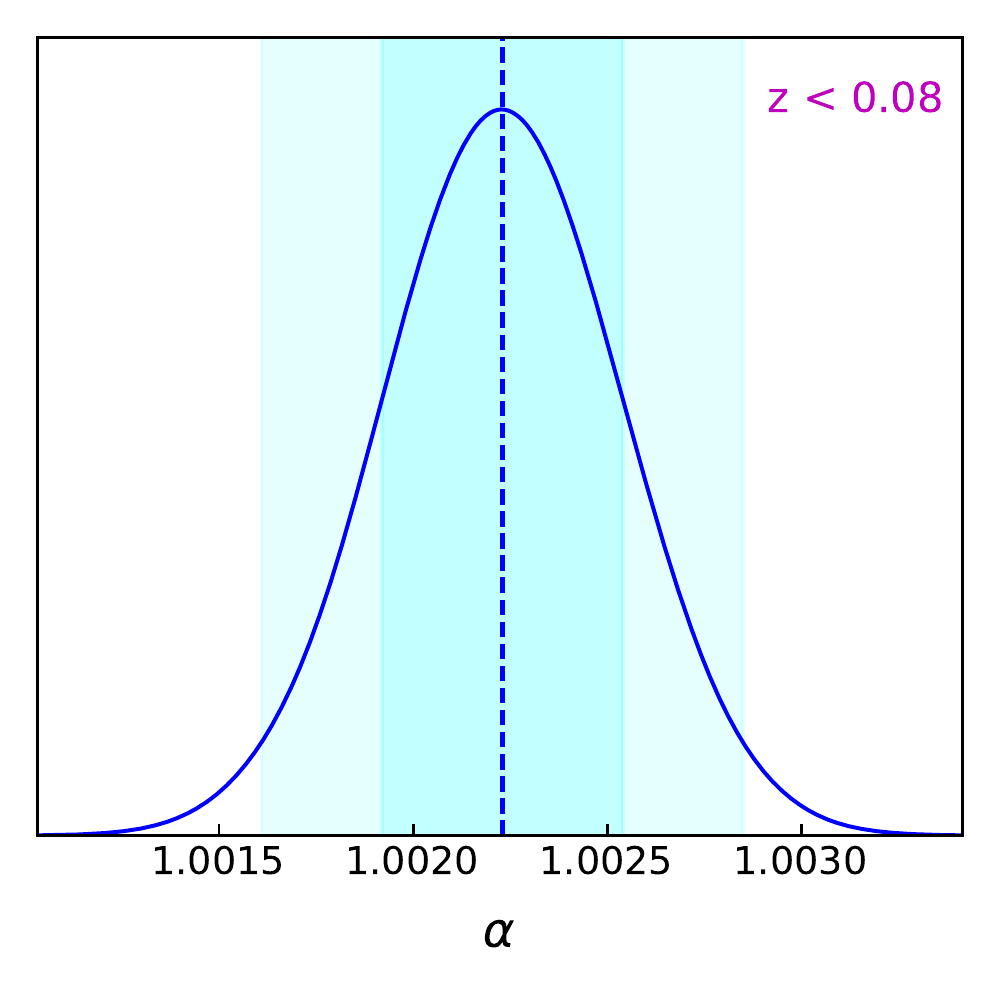}
	\includegraphics[scale=0.43]{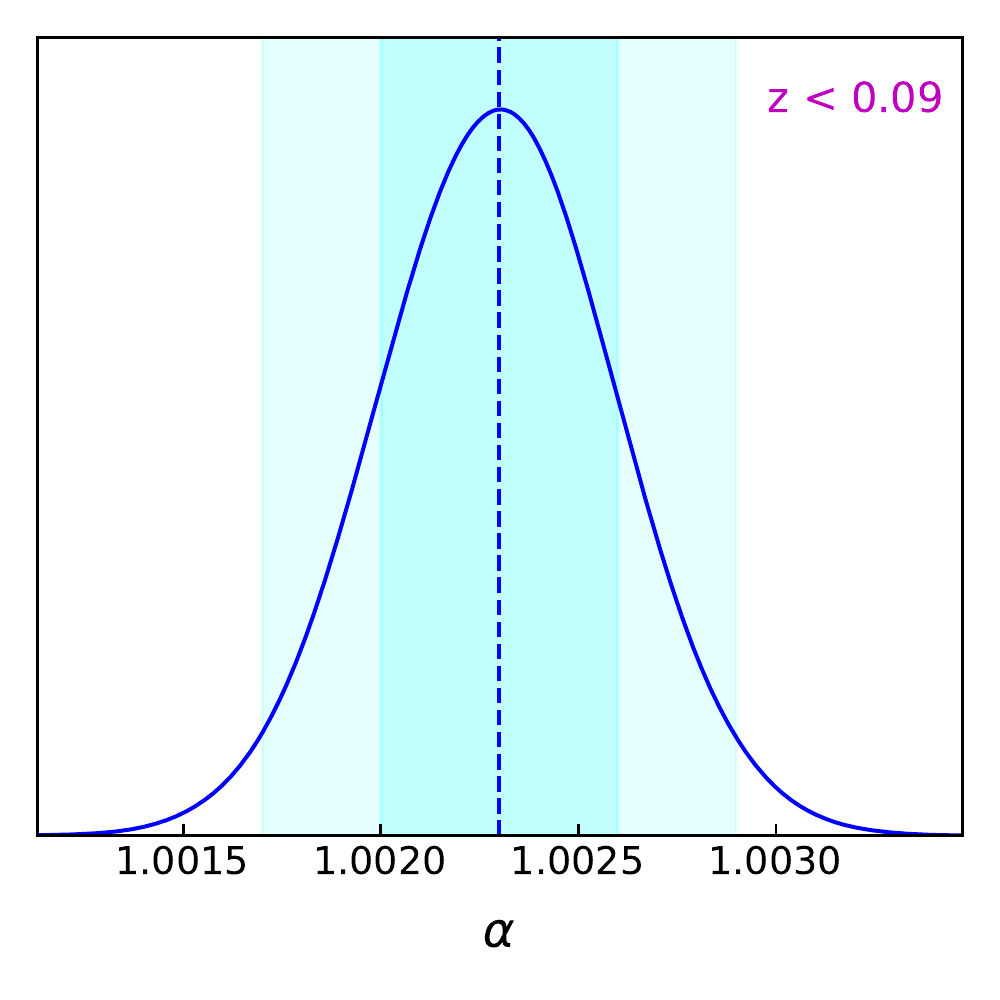}
	\includegraphics[scale=0.43]{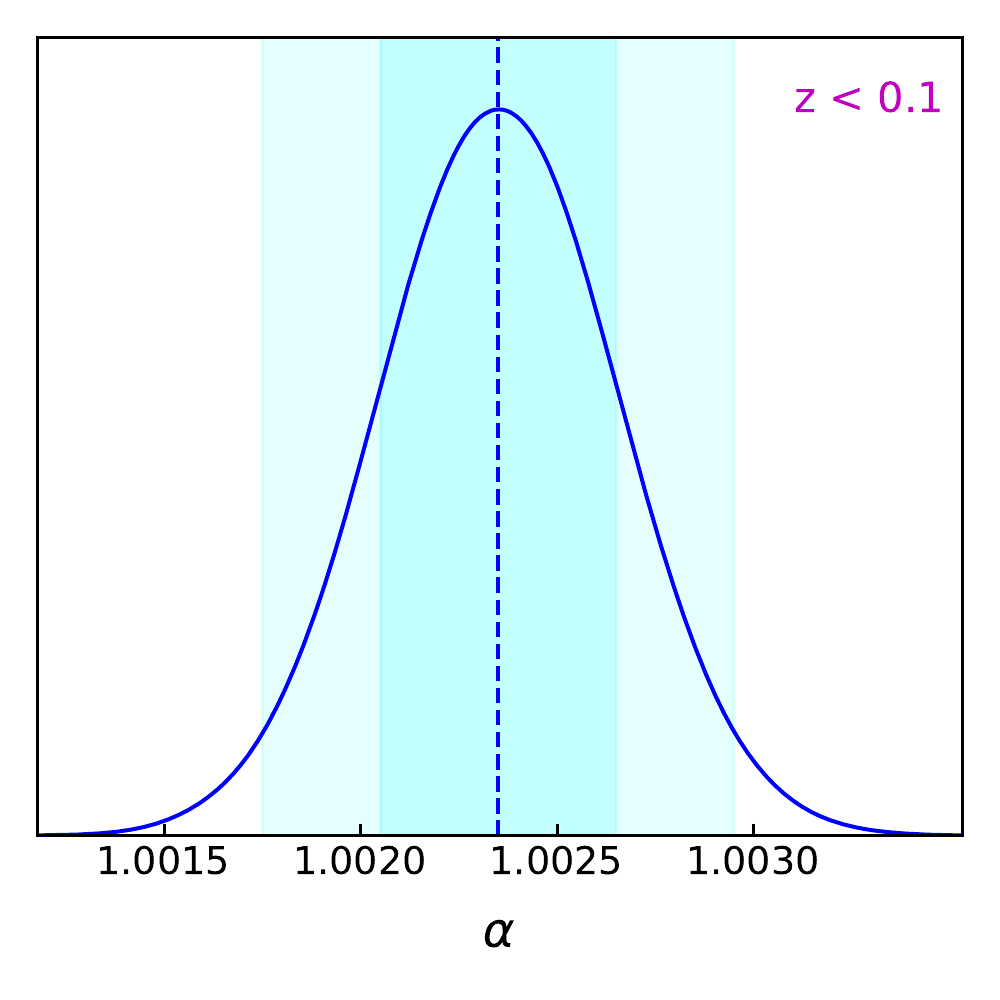}
	\includegraphics[scale=0.43]{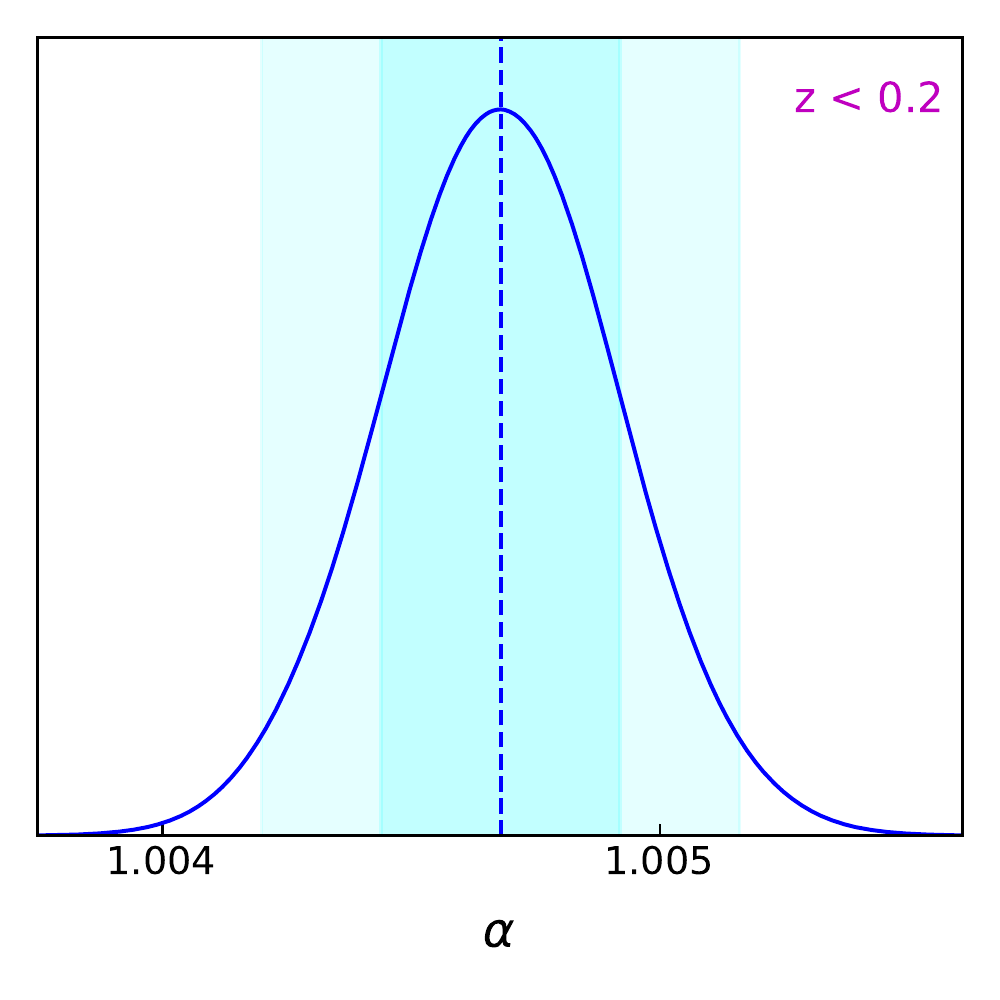}
	\includegraphics[scale=0.43]{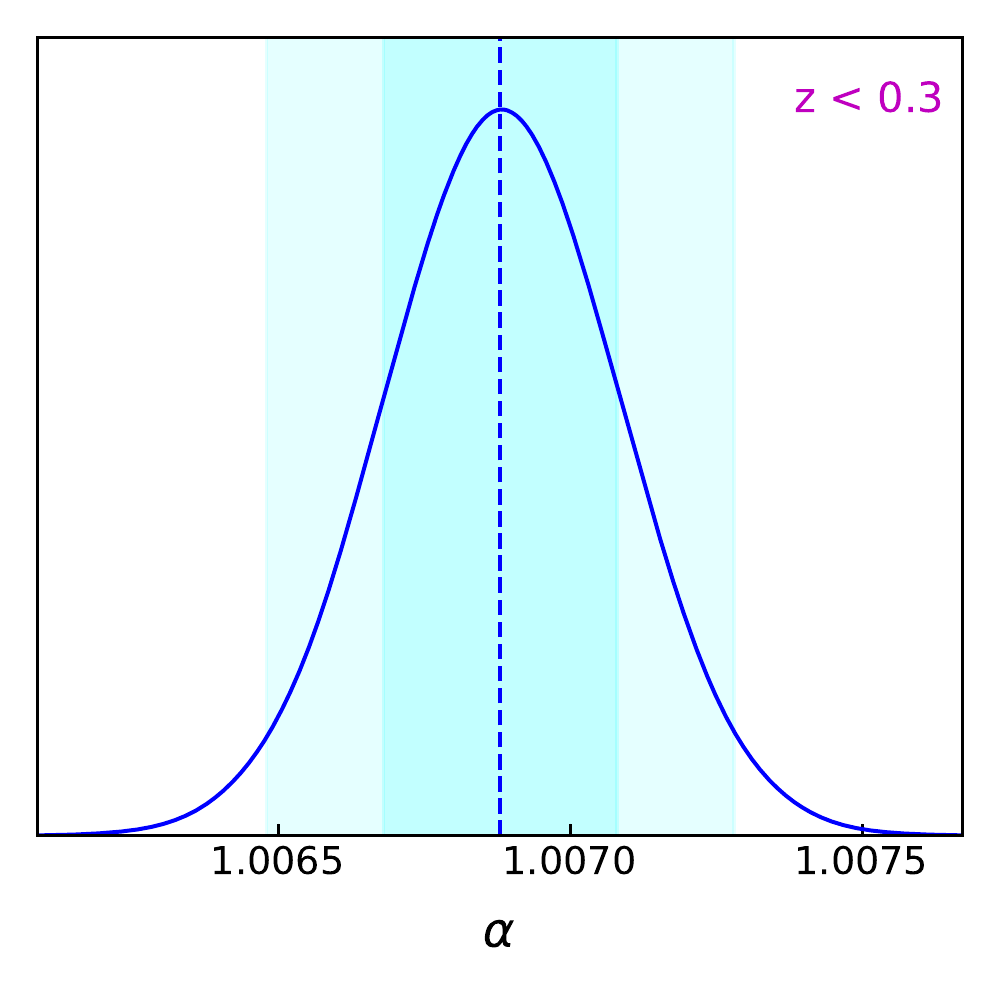}
	\caption{The 1-dimensional posterior distributions of the deviation parameter $\alpha$ in different low-z ranges of SNe Ia. The vertical bands are the $1\,\sigma$ (dark) and $2\,\sigma$ (light) confidence ranges of $\alpha$. The dashed lines denote $\alpha=1$ (red) and the best fits (blue), respectively.}\label{f3}
\end{figure*}

\begin{figure}
	\centering
	\includegraphics[scale=0.55]{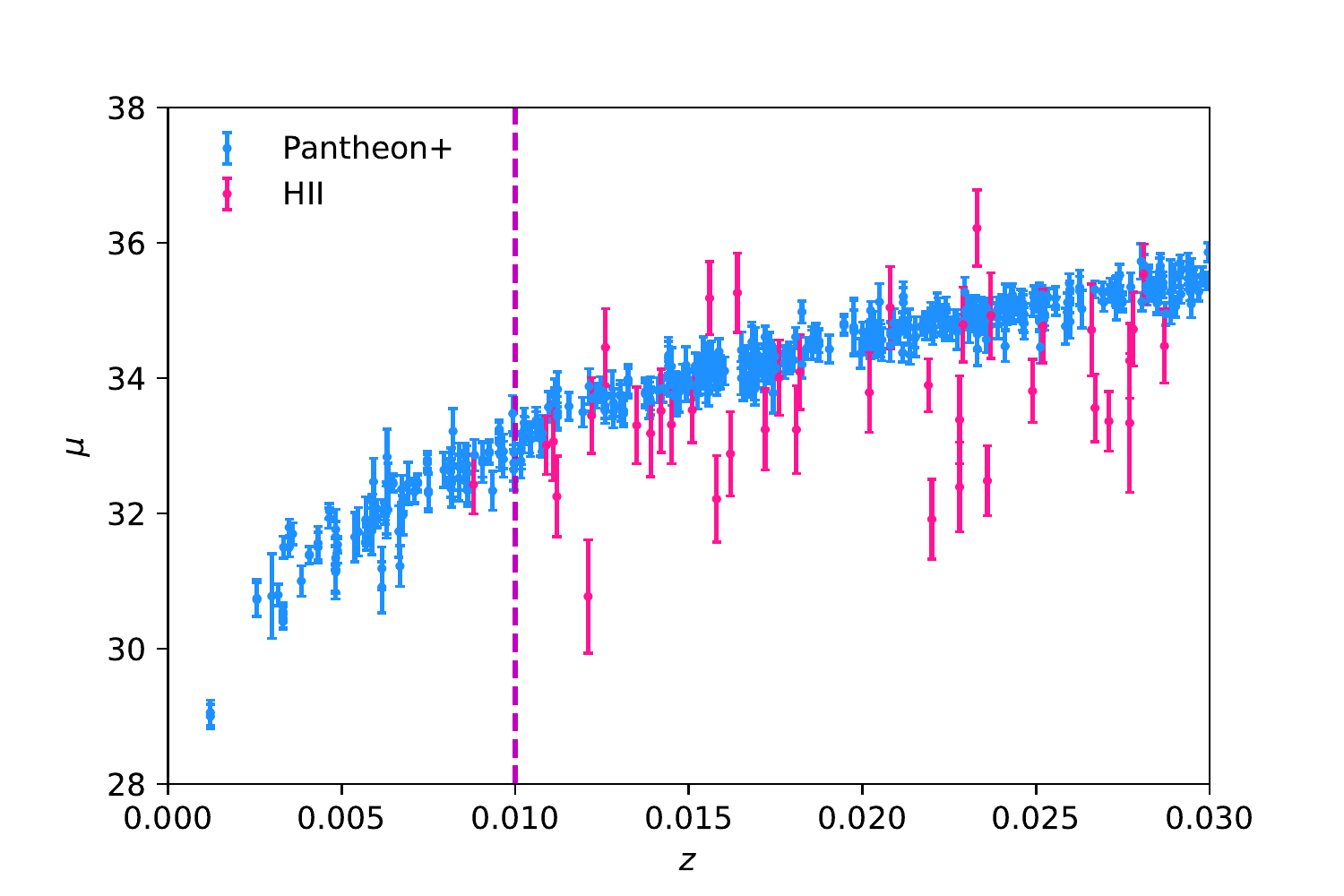}
	\caption{The redshift ($z$)-distance modulus ($\mu$) diagram of Pantheon+ (blue) and HII galaxy measurements (pink). The dashed line denotes the redshift $z=0.01$.}\label{f4}
\end{figure}

\begin{figure}
	\centering
	\includegraphics[scale=0.5]{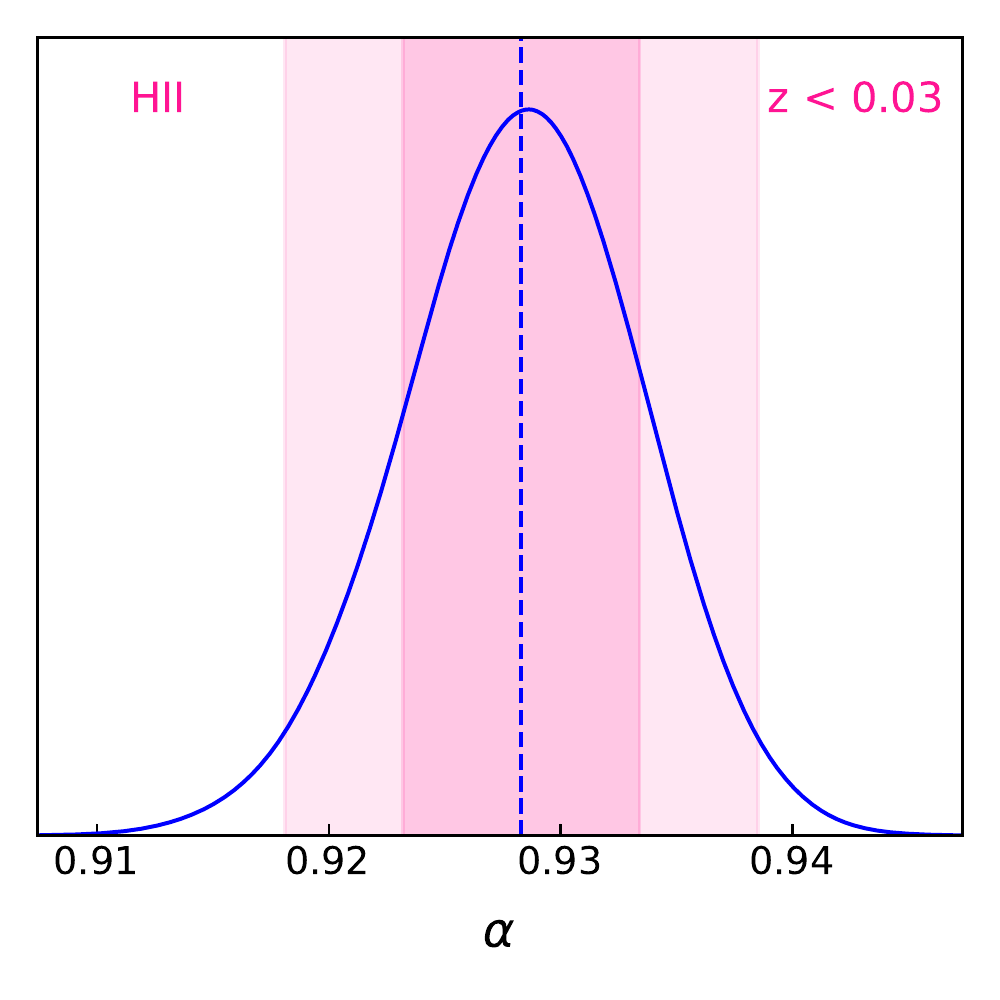}
	\caption{The 1-dimensional posterior distribution of the deviation parameter $\alpha$ for HII galaxies. The vertical bands are the $1\,\sigma$ (dark) and $2\,\sigma$ (light) confidence ranges of $\alpha$. The dashed blue line denote the best fit.}\label{f5}
\end{figure}

{\it Data and method}. Recently, Refs.\cite{Scolnic:2021amr,Brout:2022vxf} release the latest and largest SNe Ia sample called Pantheon+, which consists of 1701 light curves of 1550 spectroscopically confirmed SNe Ia coming from 18 different sky surveys. Pantheon+ has a significant increase relative to the original Pantheon \cite{Pan-STARRS1:2017jku} at low redshifts and covers the redshift range $z\in[0.00122, \, 2.26137]$. To describe the difference of two datasets better, we show the redshift-apparent magnitude diagram in Fig.\ref{f2}. Especially, Pantheon+ has 111 SNe Ia when $z<0.01$ but Pantheon does not. If considering a larger redshift range $z<0.03$, Pantheon+ has 468 points which are more than 95 SNe Ia from Pantheon by a factor of $\sim$ 5. This means that Pantheon+ can give an unprecedented constraint on the HL, and that one can test the HL at extremely low redshifts via Pantheon+ at present.

To test the HL with low-z Pantheon+ data, we propose a generalized form of HL as 
\begin{equation}
(cz)^\alpha=H_0D_L,   \label{3}
\end{equation}  
where $\alpha$ characterizes the deviation from HL and it reduces to the standard HL when $\alpha=1$. $D_L$ is expressed as 
\begin{equation}
D_L=10^{\frac{\mu-25}{5}},   \label{4}
\end{equation}  
where $\mu$ is the distance modulus of a SNe Ia where the absolute magnitude has been determined by the SH0ES Cepheid host distances \cite{Scolnic:2021amr,Brout:2022vxf}.  
To implement the Bayesian analysis, we express the $\chi^2$ at low redshifts as
\begin{equation}
\chi^2=\sum\limits_{i}\left[\frac{(cz_i)^\alpha-(H_0D_L)_{obs}}{\sigma_{H_0D_L}}\right]^2,   \label{5}
\end{equation} 
where $z_i$ is the redshift of $i$-th SNe Ia, and $(H_0D_L)_{obs}$ is our observable and it can be derived by combining the newest local measurement \cite{Riess:2021jrx} $H_0=73.04\pm1.04$ km s$^{-1}$ Mpc$^{-1}$ with the distance modulus observations from Pantheon+ \cite{Scolnic:2021amr,Brout:2022vxf}, and its error $\sigma_{H_0D_L}=\sqrt{H_0^2\sigma_{D_L}^2+D_L^2\sigma_{H_0}^2}$, where $\sigma_{H_0}$ and $\sigma_{D_L}$ denote the $H_0$ error and the uncertainty of luminosity distance of a SNe Ia at $z$, respectively.

{\it Constraints from SNe Ia}. To explore the impacts of different low-z ranges on constraints on the generalized HL in Pantheon+, we start with $z<0.01$ and increase gradually the upper limit of the range by 0.01 each time. 
The constraining results of $\alpha$ are shown in Fig.\ref{f3} and Tab.\ref{t1}. Using the unique data lying in $z<0.01$ in Pantheon+, we obtain $\alpha=1.00330\pm0.00110$, which surprisingly gives a $0.1\%$ determination on $\alpha$. Its best fitting value 1.00330 just has a $0.33\%$ difference relative to the standard HL prediction $\alpha=1$. Basically, this result means that the HL holds in $z<0.01$. Interestingly, we observe a $3\,\sigma$ inconsistency with $\alpha=1$ at such a high precision, which may indicate that the original HL needs a very small modification. When $z<0.02$, we have  $\alpha=1.00251\pm0.00068$, which is now a $0.07\%$ determination on $\alpha$ and the best fit of which has a $0.25\%$ difference relative to $\alpha=1$. It shows a $3.69\,\sigma$ inconsistency to $\alpha=1$ that is a little higher than the case of $z<0.01$ does. It is interesting that the constraint $\alpha=1.00118\pm0.00044$ for $z<0.03$ including 468 SNe Ia are just in a $2.68\,\sigma$ tension with the standard HL, while it determines $\alpha$ with a $0.04\%$ precision and its best fit has only $0.12\%$ deviation from $\alpha=1$. Interestingly, when considering $z<0.04$, the constraint $\alpha=1.00190\pm0.00036$ shows a beyond $5\,\sigma$ inconsistency to the original HL and its best fit has a larger deviation ($0.2\%$) from $\alpha=1$ than the case of $z<0.03$. Note that this is a transition point in our analysis. Staring from $z<0.04$, for the cases of $z<0.05, \, 0.06, \, 0.07, \, 0.08, \, 0.09, \, 0.1$, this discrepancy is more and more significant, the best fits prefer a positive deviation from $\alpha=1$, and the errors of $\alpha$ decrease continuously (see Tab.\ref{t1}). 

Until $z<0.1$, we find that the HL seems to describe data well in the Hubble diagram. However, when $z>0.1$, the HL exhibits a significant deviation from Pantheon+ data. To show this feature, we also constrain $\alpha$ for $z<0.2$ and $z<0.3$, and find that these two cases give $19.5\,\sigma$ and $34.4\,\sigma$ tensions. With the increase of SNe Ia number in a considered high-redshift range, the best fit of $\alpha$ will be larger, its error becomes compressed by increasing sample size, and consequently, the tension will be larger. Here we define the tension as $T\equiv\left|\alpha-1\right|/\sigma_\alpha$, where $\sigma_\alpha$ denote the error of $\alpha$ derived from observations. This statistical quantity describes the extent of the deviation from the standard HL for a given dataset. From the lower panel of Fig.\ref{f3} and Tab.\ref{t1}, it is easy to observe that $z=0.03$ can serve as a safe breaking point of HL, which is mainly ascribed to the small $T$ value (2.68) and small difference ($0.12\%$) when $z<0.03$. Hence, we choose conservatively the transition redshift $z_t=0.03$ for Pantheon+ and the accompanied transition luminosity distance $D_{L,t}=123.13\pm1.75$ Mpc. This gives the following relation     
\begin{equation}
\begin{cases}
cz=H_0D_L,     &  \quad\quad\quad z\in(0, \,\, 0.03]      \\
cz\neq H_0D_L, &  \quad\quad\quad z\in[0.03, \,\, +\infty]
\end{cases}
\end{equation}
which indicates that HL breaks down when $z>0.03$ and we should consider the cosmology-dependent recession velocity.  

{\it Constraints from HII galaxies}. In light of the above success of Pantheon+ in testing the HL, an important task is to compare the ability of alternative modern background probes with that of SNe Ia in constraining $\alpha$. This alternative probe should at least include two features: (i) it can give competitive constraints on $\Lambda$CDM relative to SNe Ia; (ii) its sample should cover $z<0.03$ with enough data points. To the best of our knowledge, HII galaxies (hereafter HII) measuring the distance via the $L(H\beta)$-$\sigma$ relation (see \cite{Gonzalez-Moran:2021drc} for details) just satisfy both requirements. Therefore, we take HII to implement a cross check and its Hubble diagram is shown in Fig.\ref{f4} for $z<0.03$. We observe that HII data not only has large errors, but also distributes in an incompact way unlike SNe Ia. This implies that HII can not give a good enough test of HL. Confronting Eq.(\ref{3}) with HII, we obtain $\alpha=0.9283\pm0.0051$ (see Fig.\ref{f5}), which gives a $0.55\%$ determination on $\alpha$, and shows, respectively, $14.06\,\sigma$ and $14.24\,\sigma$ tensions to $\alpha=1$ and Pantheon+ constraint for $z<0.03$.      

{\it Discussions and Conclusions.} Using the latest and largest Pantheon+ SNe Ia sample, we implement a modern test of the HL and confirm the linear HL with a precision of less than $0.1\%$. We find that the transition redshift $z_t=0.03$ and distance $D_{L,t}=123.13\pm1.75$ Mpc, which implies that HL holds when $z<0.03$ and breaks down at a distance of $D_L>123.13$ Mpc. 

To provide another perspective, the HL can be expressed as $D_L=cz/H_0$, which is the 1st-order Taylor expansion of $D_L$ on $z$. If calculating $D_L$ at high redshifts, one must consider the cosmology dependence. The cosmological implication of $z_t=0.03$ is that {\it any background probe can not distinguish different cosmological models when $z<0.03$.} Although the constraint from HII has a precision of less than $1\%$, it can not give an persuasive test since its incompact distribution in the late-time Hubble diagram. Consequently, we should stress the strong power of SNe Ia in exploring the low-z physics. However, except unknown systematic uncertainties, we do not rule out the possibility that the $\alpha$ tension between SNe Ia and HII predicts the existence of new physics at low redshifts. Our results are closely related to the validity of relativistic Doppler shift formula, which indicates that our constraint is degenerate with the validity of special relativity and classical Doppler shift formula at a cosmological distance.   

{\it Acknowledgements.} DW thanks Adam Riess for helpful discussions. DW is supported by the National Science Foundation of China under Grants No.11988101 and No.11851301.

\end{document}